\documentclass[a4paper,twocolumn,unpublished]{quantumarticle}
\pdfoutput=1

\usepackage[numbers,sort&compress]{natbib}
\usepackage{graphicx}
\usepackage{dcolumn}
\usepackage{amsmath}
\usepackage{amssymb}
\usepackage{float}
\usepackage[utf8]{inputenc}
\usepackage[T1]{fontenc}
\usepackage{todonotes}
\usepackage{hyperref}
\usepackage{braket}
\usepackage{siunitx}
\usepackage{lipsum}
\usepackage{mathtools}
\usepackage[dvipsnames]{xcolor}

\DeclareSIUnit{\G}{\text{G}}
\newcommand{\figref}[2]{\hyperref[#1]{Fig.~\ref*{#1}#2}} 

\newcommand{\tDelta}{\tilde\Delta}

\colorlet{mylinkcolor}{RoyalPurple}
\colorlet{mycitecolor}{RoyalPurple}
\colorlet{myurlcolor}{RoyalPurple}
\hypersetup{
	linkcolor  = mylinkcolor,
	citecolor  = mycitecolor,
	urlcolor   = myurlcolor,
	colorlinks = true,
	breaklinks = true
}

\newcommand{\subfig}[1]{(#1)}
\newcommand{\dd}[2]{\frac{\partial #1}{\partial #2}}
\newcommand{\iu}{{i\mkern1mu}}
\newcommand{\fuw}{\affiliation{Faculty of Physics, University of Warsaw, Pasteura 5, 02-093 Warsaw, Poland}}
\newcommand{\cent}{\affiliation{Centre for Quantum Optical Technologies, Centre of New Technologies, University of Warsaw, Banacha 2c, 02-097 Warsaw, Poland}}

\renewcommand{\paragraph}[1]{}

\begin{document}
\title{Multiplexed storage and interaction of Rydberg spinwaves via the gradient echo memory protocol}

\author{Bartosz Niewelt}
\thanks{Equal contributions}
\fuw\cent
\author{Stanisław Kurzyna}
\thanks{Equal contributions}
\fuw\cent
\author{Bartosz Kasza}\fuw\cent
\author{Wojciech Wasilewski}\fuw\cent
\author{Michał Parniak}\fuw\cent
\email{mparniak@fuw.edu.pl}

\maketitle
\begin{abstract}
Collective Rydberg excitations offer strong and controllable interactions for quantum information processing, sensing, and nonlinear quantum optics, but their integration with temporally or spectrally multiplexed schemes, such as the Gradient Echo Memory (GEM) protocol, is hindered by rapid motional dephasing caused by the large spinwave wavevector. We demonstrate a new type of multi-photon addressing and interfacing scheme (with levels following the shape of the letter \'{N}) that allows us to generate collective Rydberg excitations with near-zero momentum transfer, extending the Rydberg spinwave lifetime almost tenfold. The scheme relies on two additional off-resonant driving fields arranged at a magic angle, forming a closed wavevector loop while remaining compatible with GEM-induced inhomogeneous broadening. This enables storage and manipulation of long-lived Rydberg spinwaves in a multimode quantum memory. Using microwave coupling between neighboring Rydberg states, we can control the attenuation between stored excitation modes by interaction-induced decay and demonstrate interaction-controlled diffraction of a retrieved optical signal. Our results reestablish compatibility between Rydberg excitations and GEM, providing a route toward multimode quantum memories with controllable long-range interactions and applications in quantum networking, sensing, and quantum information processing.
\end{abstract}

\section{Introduction}

Neutral atoms excited to high-lying Rydberg states constitute a powerful platform for quantum technologies due to their strong and controllable interactions. A hallmark feature of these systems is the Rydberg blockade\cite{Urban2009,Gaëtan2009}, which enables the implementation of high-fidelity quantum gates\cite{PhysRevLett.104.010503,PhysRevLett.123.170503,Evered2023,PhysRevLett.112.040501} and provides a viable route toward scalable, fault-tolerant quantum computation\cite{Bluvstein2024}.

These strong interactions have also enabled a wide range of applications in quantum optics. Rydberg-mediated nonlinearities facilitate photon–photon interactions\cite{Chang2014,Firstenberg2013}, single excitation generation\cite{PhysRevLett.117.180501}, and study of effects of photon-atom entanglement\cite{Li2013}. Rydberg platforms provide a controllable non-linear medium\cite{PhysRevLett.105.193603, Gorniaczyk2016} holding potential as a deterministic single photon source\cite{doi:10.1126/science.1217901}.

At the same time, Rydberg systems serve as a versatile platform for quantum simulation, where long-range interactions play a key role in exploring spin chain evolution\cite{PhysRevLett.120.113602,PhysRevLett.114.113002,PhysRevX.8.011032,Scholl2021}, quantum phase transitions\cite{Keesling2019}, and in the study of many-body dynamics\cite{Schauß2012}. 

Beyond interaction strength, Rydberg atoms exhibit extreme sensitivity to external fields due to their large dipole moments on transitions to neighboring states. This property has been exploited for qubit manipulation\cite{PhysRevLett.127.063604}, measurement of thermal black-body background\cite{Borówka2024}, and precision sensing across a broad range of the electromagnetic spectrum\cite{jlrg-6889,Nowosielski:24,PhysRevApplied.22.034067}. Other applications, such as microwave/mmWave-to-optical conversion\cite{Borówka2024,PhysRevA.99.023832,Kumar2023}, enable optical detection and measurement for hybrid quantum sensors.

\paragraph{Collective excitations}
The light–matter coupling is further enhanced by collective excitations in atomic ensembles\cite{PhysRevLett.99.163601,Murray_2016}, 
taking advantage of multiple correlated quantum emitters. One of the most prominent applications are quantum memories, where a quantum state of light is mapped onto delocalized atomic coherence and retrieved on demand\cite{PhysRevA.98.042126}. This makes them a crucial component in long-distance quantum communication and quantum repeater architectures\cite{Duan2001, PRXQuantum.2.040307}.

\paragraph{GEM}

One of the techniques that allows spectral engineering of the atomic coherence is Gradient Echo Memory (GEM)\cite{Buchler:10,Sparkes_2013}. By introducing a controlled inhomogeneous broadening of the energy levels, GEM enables the different Fourier components of probe light to be mapped spatially and later manipulated. Spatial and temporal multiplexing capabilities\cite{Hosseini2011,PhysRevA.86.023801,Lipka_2021,Parniak2017} enable multiple independent modes to be stored simultaneously. Methods such as Spatial Spinwave Modulation\cite{PhysRevApplied.11.034049,PhysRevLett.122.063604} have proven to expand the capabilities of GEM to also modify the stored atomic coherence. This sparked a wave of new applications, including superresolution\cite{Mazelanik2022}, programmable spatial dispersion\cite{PhysRevA.109.012418} and fractional Fourier transform\cite{PhysRevLett.130.240801}. 
A hybrid implementation of GEM and Electromagnetically Induced Transparency (EIT) was also studied recently\cite{Papneja:26,8wzh-jk9z}.

The proposed scheme for the creation of the long-lived atomic coherence can be effectively combined with the GEM mode multiplexing capability. 
Multiple subsequent pulses can be stored in the same atomic ensemble as Rydberg spinwaves with different wavevectors.
Moreover, frequency-to-position mapping in GEM allows for controllable spatial distribution of the Rydberg spinwaves, enabling switching the position-dependent strong Rydberg-Rydberg dipolar interactions with microwave (MW) pulses. 

\paragraph{Motional dephasing}
However, integrating Rydberg excitations with GEM remains an outstanding challenge. The primary obstacle in merging the two technologies is the short lifetime due to the motional dephasing \cite{z6dc-zrnh,Li2016,PhysRevA.98.033411,PhysRevLett.127.063604}, 
arising from typically large wavevector of the collective Rydberg excitation. In other words, the phase of the Rydberg component rapidly changes along the atomic memory and is thus vulnerable to movement of atoms on a micrometer scale. The residual thermal motion of the atoms amounts to phase decoherence of Rydberg spinwaves 
and quickly degrades the efficiency of readout from the memory. 

\paragraph{Mitigation w/o GEM}
Recently, several techniques have been developed to mitigate the effects of atomic motion. 
One approach is based on trapping atoms in a magic-wavelength optical lattice\cite{PhysRevA.98.033411}, stopping the movement entirely. 
Numerous methods rely on pulsed protocols, including  spatial phase modulation\cite{Kurzyna2024longlivedcollective} and wavevector cancellation by transfering either the Rydberg state coherence \cite{z6dc-zrnh,PhysRevLett.134.053604}, or the ground state coherence\cite{PhysRevA.93.063819,kurzyna2025microwavefieldquantummetrologyinherent} using additional beams positioned at a specific angle. 

\paragraph{Problem statement}
Motional decoherence itself excludes few-\unit{\us} long GEM storage 
and the stored information dephases before the echo signal can be read out.
Thus pulsed techniques cannot be applied.

\paragraph{Solution}
Here, we present a novel four-photon light-atom interface in a \'{N} letter shape configuration
that couples light to nearly zero wavevector component of Rydberg atoms
thus re-enabling GEM and prolonging lifetime. 
To the typical counterpropagating probe-coupling interface, we 
add two driving beams crossed on the atomic ensemble at a "magic angle".
The wavevectors of all engaged beams in the process form a closed figure. 
The added beams drive virtual transitions on the D2-line, thus enabling best power efficiency. 
The spinwave stored  this way has a zero wavevector, making it almost entirely insensitive to the motional dephasing. 
The interface is off-resonant, allowing it to be applied alongside GEM inhomogeneous broadening of the energy levels in the medium, enabling the control of the spatial distribution of the Rydberg dipole interactions and controlled interaction switching.

\section{Four-photon Ń configuration}
\begin{figure}
    \centering
    \includegraphics[width=\columnwidth]{ 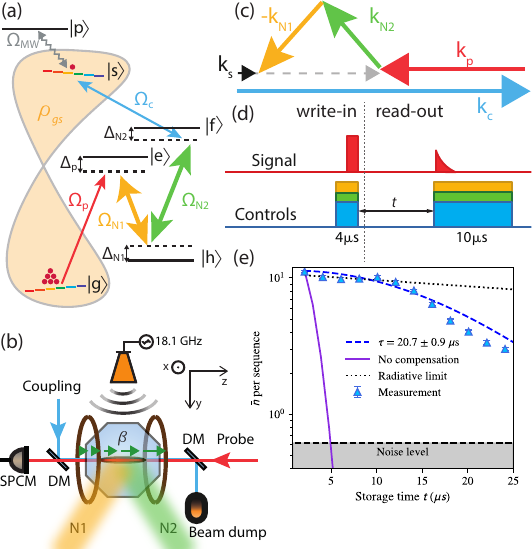}
    \caption{
            \subfig{a} Energy levels structure of $^{87}$Rb relevant in the experiment for the Rydberg GEM interface, 
            \subfig{b} Simplified scheme of the experimental setup for storing Rydberg coherence in the GEM.
            \subfig{c} Geometrical representation of the beams' wavevectors used to instantaneously create the spinwave with reduced wavevector. Grey dashed arrow represents uncompensated spinwave wavevector $k_\text{s}$ in a typical ladder-type scheme.
            \subfig{d} Simplified sequence for the lifetime extension protocol. All of the controls (N1, N2, coupling) are enabled at the same time.
            \subfig{e} Experimental data for the lifetime extension protocol. Blue triangles represent the number of readout photons collected with SPCM, while the dashed line corresponds to the Gaussian decay fitted to experimental data. The solid line represents the Rydberg spinwave decay without the lifetime extension protocol, with thermally-limited decay $\tau_{\text{th}} = \SI{2.3}{\micro \s}$.}
    \label{fig:levels_scheme_lifetime}
\end{figure}

\paragraph{Motional decoherence}
Typical two-photon atom-light interface in a ladder configuration
uses the coupling laser to interface the probe beam 
to the Rydberg coherence. 
The phase of the ground-Rydberg coherence 
arises from both beam phases $\exp(i k_\text{p} z)$ and $\exp(i k_\text{c} z)$ 
with $k_\text{p}$ and $k_\text{c}$ denoting their wavevectors.
For a favourable, counterpropagating configuration, 
said phase varies quickly along $z$ as $ \propto \exp[ i (k_\text{c}-k_\text{p}) z] = \exp (i k_\text{s} z)$.
In other words, the created atomic coherence has a large wavevector $k_\text{s} = k_\text{c} - k_\text{p}$.

The thermal motion of the atoms with average velocity $v_\text{th} = \sqrt{k_B T / m}$, 
causes phase decoherence, blurring the spin waves on the microsecond timescale. 
The motional limit of efficiency follows a Gaussian function $\eta_\text{th}(t) = \exp\left[-(t/\tau_\text{th})^2\right]$ with $\tau_\text{th} = (v_\text{th} k_\text{s})^{-1} \approx \qty{2}{\micro \s}$
for $\SI{780}{\nano \meter}$ and $\SI{480}{\nano \meter}$ beams used in Rubidium.

\paragraph{Ń configuration}

To overcome the motional decoherence and increase the lifetime, 
we engineered a configuration of lasers that couples the signal beam to a Rydberg spinwave with the reduced wavevector $\boldsymbol{k}_\textbf{s}\approx 0$.
This is enabled by employing four beams in the energy configuration presented in \figref{fig:levels_scheme_lifetime}{a}. 
The two control beams N1 and N2 are crossed at an angle $\alpha$ to the $z$-axis. 
This way, the wavevectors of light fields involved can form a closed loop as shown in \figref{fig:levels_scheme_lifetime}{c}. 
The resulting wavevector of the spinwave created in the four-photon process equals:
\begin{equation}\label{eq:wavevectors}
\boldsymbol{k}_\textbf{s} = \boldsymbol{k}_{\textbf{c}} + \boldsymbol{k}_\textbf{p} - \boldsymbol{k}_\textbf{N1} + \boldsymbol{k}_\textbf{N2} \approx 0,
\end{equation}
where the sign of the $\boldsymbol{k}_\textbf{N1}$ is reversed because it drives a stimulated de-excitement.

\paragraph{Other configurations}
In general, engineering the light-atom interface with adjustable momentum transfer requires at least three light beams. 
Three-step ladder configurations of this type have been proposed\cite{qzwk-2g33,PhysRevApplied.20.L061004}. 
However, they require that the driving lasers for weak transitions be at an angle to the atomic ensemble. 
When the atomic ensemble is long, this means distributing available power over a large surface area and prohibits high Rabi frequencies.
In our scheme, we purposely chose the off-axis lasers to couple to the strongest transition. 
This way, the large Rabi frequencies can be reached with reasonable power even with nearly perpendicular beams. 

Let us also discuss the viability of the continuous dressing\cite{PhysRevX.11.011008}, where the collective excitation is coupled to a state experiencing an opposite motional shift. The reduction of much larger wavevector mismatch would require substantially more powerful lasers to produce the required light shifts for the larger ensemble ($\approx 6000$ more power).

\subsection{Simulations}
Numerical simulations for the \'{N}-GEM are performed by solving the optical Bloch equations for a propagating probe pulse, which we describe by its Rabi frequency $\Omega_p(z,t)$. The probe is coupled to atomic coherence $\rho_{gs}(z,t)$ between ground state $\ket{g}$ and Rydberg state $\ket{s}$ as depicted in \figref{fig:levels_scheme_lifetime}{a}. The joint evolution is found by solving a pair of propagation equations:
\begin{equation}
    \begin{cases}
    \begin{aligned}
    &\dd{}{z}\Omega_p = \frac{n(z)}{N} \, \frac{\text{OD}}{2} \,\left[ C_{\Omega,\Omega} \Omega_p + C_{\Omega,\rho} \rho_{gs}\right] \\[1ex] 
    &\dd{}{t}\rho_{gs} = C_{\rho,\Omega} \Omega_p + \left[C_{\rho,\rho} + \iu \delta(z) - \frac{\gamma_{rr}}{2} \right]   \rho_{gs} 
    \end{aligned}
    \end{cases}
    \label{MBEs}
\end{equation}
where $n(z)$ is the atomic density profile, OD is the optical depth, $N$ is the number of atoms in the ensemble, and $\delta(z)$ is position-dependent detuning,
$C_{i,j}$ are complex coupling coefficients and $\gamma_{rr}$ is the interaction-induced decoherence. 
The coefficients $C_{i,j}$ depend on laser detunings $\Delta_p,$ $\Delta_\text{N1}$, $\Delta_\text{N2}$ and Rabi frequencies $\Omega_c$, $\Omega_\text{N1}$, $\Omega_\text{N2}$
and are obtained from the Hamiltonian of the five-level atom described in the Appendix \autoref{sec:theory}. 

\paragraph{GEM mapping}
The magnetic field gradient in the GEM scheme causes Zeeman shifts in the energy levels between $\ket{g}$ and $\ket{s}$ states. In the experiment, a GEM field with gradient $\beta = \qty{0.47}{\G \per \cm}$ is applied, the total memory bandwidth is $B = 2\pi\times\qty{0.6}{\mega \hertz}$.
This maps the different spectral components of the signal light to positions on the atomic ensemble along the $z$-axis, with the detuning given by $\delta(z) = \beta z$. 
Thus, the induced position-dependent coherence $\rho_{gs}(z)$ is proportional to the Fourier transform of the envelope of the write signal 
$\tilde\Omega_p(\delta)=\mathcal F_{t\rightarrow \delta}[\Omega_p(t)]$. 
The temporal multiplexing of the incoming signal photons is achieved by mapping the complete shape of the signal $\Omega_p(t)$ 
to the components of the coherence in the wavevector domain $\tilde \rho_{gs}(k_z)=\mathcal F_{z\rightarrow k_z}[\rho_{gs}(z)]$. This is possible because of continuous imprinting of the magnetic phase $\exp (i \beta t z)$\cite{hosseini_coherent_2009}.
After writing the optical pulses, the gradient of the magnetic field is reversed, and after rephasing, the signal can be read out. In the experiment, the reversing and rephasing take about \qty{16.9}{\us}.

\paragraph{Interactions}
The Rydberg-Rydberg interactions between $\ket{s}$ and $\ket{p}$ were modeled by adding a decay $\gamma_{rr}$ of the coherence $\rho_{gs}$, which is proportional to the interaction strength between Rydberg states $C_3$ and to the population in the state $\ket{p}$ \cite{kurzyna2025microwavefieldquantummetrologyinherent}. We are working in the low-excitation regime, so the population $\rho_{pp}$ can be approximated with the square of the coherence $|\rho_{gp}|^2$.
The decay mechanism is described in more detail in the Appendix \autoref{sec:Decay}. 

\paragraph{MW rotation}
In addition to the above equations, the action of a short MW pulse can be modelled as an instantaneous rotation 
$U(t)=\exp(\iu\sigma^{(sp)}_x \Omega_\text{MW}t/2 )$ between states $\ket{s}$ and $\ket{p}$.

\section{Experimental setup}
The experiment uses a quantum memory based on $^{87}$Rb atoms trapped in a magneto-optical trap. The setup is presented in \figref{fig:levels_scheme_lifetime}{b}. The atomic cloud is cigar-shaped along $z$ direction, and measures (FWHM) $\qty{1.2}{\mm}\times\qty{1.2}{\mm}\times\qty{5}{\mm}$. The atoms are cooled down to a temperature of $\qty{80}{\micro\K}$. 
The experiment is performed in a sequence with $\qty{200}{\Hz}$ repetition rate, which includes trapping and pumping. 

\paragraph{Preparation} 
MOT coils and cooling beams are turned off for the experimental sequence, which is conducted in the bias magnetic field $B = \qty{0.85}{\G}$. The magnetic field is directed along the quantization axis $z$.
After the trapping and cooling procedure, we pump the atoms to $\ket{g} = \ket{5^2 S_{1/2}\; F=1\; m_F = 1}$.
The optical depth measured at the probe laser transition 
is $\num{33}$, which corresponds to atom number $\num{2.2e7}$ and mean atomic density $\qty{4.5e10}{\per\cubic\cm}$. For simulations, we chose the spatial distribution of the ensemble $n(z)$ to have a super-Gaussian(p=4) profile with the width $\text{FWHM} = \qty{5}{\mm}$, which corresponds to the experimentally measured length of the cloud.

\paragraph{Four-photon transition}
We engineer a four-photon interface between the probe light and atomic coherence between the ground state 
$\ket{g}$ 
and the Rydberg state $\ket{s} = \ket{59^2 S_{1/2}\; m_J=1/2}$. 
The four-photon transition depicted in \figref{fig:levels_scheme_lifetime}{a} involves probe, N1, N2, and coupling lasers.
The $\sigma^+$-polarized probe light acts on the transition between $\ket{g}$ and a virtual state detuned by $\Delta_p=\qty{-30}{\MHz}$ from $\ket{e} = \ket{5^2 P_{3/2}\; F=2\; m_F = 2}$. Probe couter-propagates along the $z$-axis and has a wavevector $\boldsymbol{k}_\textbf{p} = -k_\text{p} \hat{z}$.
The N1 laser with $\pi$ polarization and Rabi frequency $\Omega_{N1}=2\pi\times\qty{25.5}{\MHz}$ continues the transition to a virtual state detuned by $\Delta_{N1}=\qty{11}{\MHz}$ from $\ket{h} =\ket{5^2 S_{1/2}\; F=2\; m_F = 2}$ state. The N2 coupling laser with $\sigma^+$ polarization and Rabi frequency  $\Omega_{N2}=2\pi\times\qty{20}{\MHz}$ links the previous excitation to a virtual state detuned by $\Delta_{N2}=\qty{20}{\MHz}$ from $\ket{f} =\ket{5^2 P_{3/2}\; F=3\; m_F = 3}$ state. The wavevectors are respectively $\boldsymbol{k}_\textbf{N1}$ and $\boldsymbol{k}_\textbf{N2}$.
Finally, the coupling laser with $\sigma^-$ polarization and Rabi frequency $\Omega_c=2\pi\times\qty{1.8}{\MHz}$ completes the transition and couples to the $\ket{s}$ state with the position-dependent four-photon detuning $\delta(z)$. Coupling propagates along the $z$-axis and has a wavevector $\boldsymbol{k}_\textbf{c} = k_\text{c} \hat{z}$. During the experiment, all the control beams illuminate the atoms simultaneously during write-in, allowing the probe to couple directly to a zero-momentum spinwave $k_\text{s}\approx 0$.
The same four-photon transition is applied to convert the spinwave back to photons.

\paragraph{Spatial alignment}
Probe and coupling propagate in opposite directions along the ensemble with waists set to $\qty{180}{\um}$ and $\qty{300}{\um}$ respectively.
To satisfy the closed-loop wavevector relation \ref{eq:wavevectors}, the N1 and N2 coupling lasers are crossed at an angle $\alpha \approx \qty{36.4}{\degree}$ in the atomic ensemble. 

Both N1 and N2 beams are collimated in the transverse direction $x$ and have waist \qty{990}{\um}. In the $y-z$ plane, the beams are diverging, and each has a width of $\qty{6.07}{\mm}$ in the cloud.
We adjust the polarizations of N1 and N2 beams to preferentially drive the desired transitions (i.e. change $m_F$ by 0 and 1, respectively). In the final alignment step, these polarizations are tuned to optimize the readout strength.

\paragraph{MW transitions}
The highly excited Rydberg state $\ket{s}$ strongly couples to a neighboring state $\ket{p} = \ket{59^2 P_{3/2} \; m_J = 3/2}$ via a dipole MW transition with frequency $f=\qty{18.1}{\GHz}$. We used a custom-made stub antenna placed at the side of the vacuum chamber to emit MW pulses with $x$ polarization, enabling the rotation between the $\ket{s}$ and $\ket{p}$ states with the Rabi frequency $\Omega_{\text{MW}} = 2\pi \times \qty{3.2}{\mega \hertz}$.

\paragraph{Measurement}
The read-out pulses pass through an optical bandpass filter and are coupled to a fiber-connected single-photon counting module SPCM-AQRH-14 Excelitas with a quantum efficiency of about 60\%. 
The measurements presented are an average of multiple repetitions of the experiment.

\section{Results}
\begin{figure*}[t]
    \centering
    \includegraphics[width=\linewidth]{ 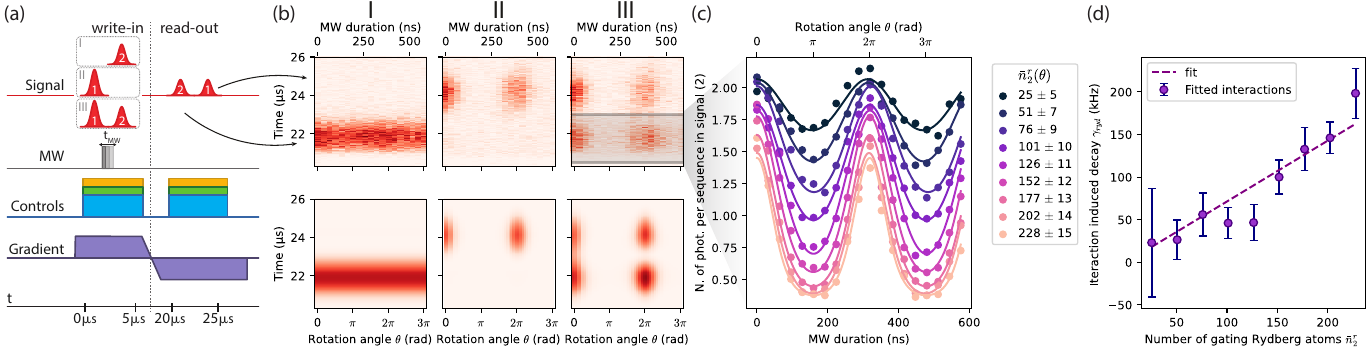}
    \caption{Controlled attenuation of the readout for different rotation angles $\theta$ and different numbers of control excitations $\bar n_1$.
    \subfig{a} Simplified sequence for the Rydberg interaction-controlled Rabi oscillation. 
    \subfig{b} Upper row: experimental data for the $\bar{n}_2^r = \qty{228}{}$ initial number of Rydberg excitations, lower row: Numerical simulations. 
    The columns correspond to
    I: only the signal pulse (2nd) is stored;
    II: only the gating pulse (1st) is stored and, by means of MW rotation, is distributed between states $\ket{s}$ and $\ket{p}$ in the ratio $\cos^2\theta$ to $\sin^2\theta$. 
    III: both pulses are stored. 
    Due to the dipolar Rydberg interactions, the atoms in $\ket{p}$ induce dephasing of the stored signal pulse. 
    \subfig{c} 
    The number of photons retrieved from the signal pulse $\bar n_2^r(\theta)$
    fitted with 
    $\bar n_2^r(0) \exp\left[-g_{rr} \bar n_1t_{\text{int}} \sin^2\theta\right]$.
    \subfig{d} 
    Interaction decay rate as a function of the total number of Rydberg atoms produced by the gate pulse $\bar n_1$.
    }
    \label{fig:MW_pulses}
\end{figure*}

\subsection{Lifetime measurement}
We benchmark the excitation scheme by measuring the lifetime of the stored coherence between states $\ket{g} \leftrightarrow \ket{s}$. For this part of the experiment, we store the light in the memory without the magnetic field gradient. The simplified experimental sequence is shown in \figref{fig:levels_scheme_lifetime}{d}, and the measurement results are presented in \figref{fig:levels_scheme_lifetime}{e}. The photons for the data points were collected for 500 sequences and averaged to obtain the mean number of detected photons for each storage time. 
To the data, we fitted a thermal-limited readout efficiency function $\bar n(t) = \bar n_0 \exp\left[-(t/\tau)^2\right]$ and obtained the extended memory lifetime $\tau = \qty{20.7(0.9)}{\us}$. For comparison, we also plotted the lifetime for the Rydberg coherence stored in the memory in a typical two-photon scheme, where the lifetime is  $\tau_{\text{th}} = \qty{2.3}{\us}$.
The memory lifetime is ultimately limited by the spontaneous emission of the Rydberg state at room temperature, in our case $\tau_\text{rad} = \qty{96}{\micro \s}$. 
This limit is depicted in \figref{fig:levels_scheme_lifetime}{e} with a dotted line. 

The acceleration in decay rate for longer storage times is due to small misalignment of N1 and N2 beams, resulting in some residual wavevector of the spinwave $k_\text{s}$.

\subsection{Controlled attenuation gate}
Temporal and spatial multiplexing enabled by the GEM allows us to probe the interactions between different highly-excited states $\ket{s}$ and $\ket{p}$. 
This is achieved by storing two probe pulses in two overlapping spinwaves with slightly different $k_\text{s}$ as shown in \figref{fig:MW_pulses}{a}.
Between the probe pulses, we apply a MW rotation so that the first pulse creates a controlled number of excitations in the $\ket{p}$ state 
$\bar n_p=\bar n_1 \sin^2\theta$ where $\bar n_1$ is the total number of excitations stored by the gate pulse (1st pulse) and $\theta$ is the rotation angle.
Specifically, we elected to control the angle of MW rotation $\theta$ by changing the MW pulse duration $t_\text{MW}$ from \qty{0}{\ns} to \qty{576}{\ns}, 
while the MW intensity was kept constant thus yielding $\theta = \Omega_\text{MW} t_\text{MW} / 2$.
The second pulse is mapped onto $\rho_{gs}$ coherence. 
The interaction happens mainly during the gradient reversal which takes $\qty{14}{\micro \s}$ 
and the sequence is finished by reading out from the memory. Note that the pulses are read out in the reverse order, i.e., the second stored pulse comes out first.

As a cross-check, we perform the experiment in three configurations: each pulse alone (I and II) and both (III). 
The measured readout pulses and the simulations for all configurations are presented in the upper and lower row of \figref{fig:MW_pulses}{b}, respectively. 

Configuration I represents the standard storage of the signal pulse (2nd), all excitations are in state $\ket{s}$. 
We verify that the pulse is not affected by the MW field by measuring the readout. 

In configuration II, we stored only the gating pulse (1st), followed by MW pulse. 
Changing the duration of the MW pulse leads to Rabi oscillations between $\ket{s}$ and $\ket{p}$, thus transfering the remaining $\rho_{gs}$ coherence to the $\rho_{gp}$ coherence.
Since the rotation is to a good extent reversible, we infer that the remainder of the populated Rydberg excitation in $\ket{s}$ was transferred to $\ket{p}$.
To characterize the attenuation efficacy, we estimate the total number of excitations stored by the gate pulse $\bar n_1$ 
from the multiphoton probe absorption, by dividing the number of readout photons by storage efficiency. The number of Rydberg excitations created with the gating impulse allows to verify the experimentally measured interaction-induced decay with the theoretically predicted value for the single excitation.

For configuration III, we store both pulses separated by $\qty{2.2}{\us}$ with the MW pulse in between. 
During the readout, we observe that the signal pulse (1st) follows the Rabi oscillation of the gating pulse (2nd). 
This is due to the interactions between different Rydberg spinwaves, which induce decay of the signal coherence $\rho_{gs}$ 
that depends on the population of the $\ket{p}$ state. 
We can modify the decay rate by changing the intensity of the gating pulse (2nd). 
The total number of detected photons from the signal pulse is presented in \figref{fig:MW_pulses}{c}. 
We repeat the sequences for a range of $\theta$ and various values of the excitation stored gate pulse $\bar n_1$ varied by adjusting the gate pulse intensity.
The decay due to interactions is confirmed by fitting the number of photons retrieved from the 2nd pulse 
$\bar n^r_{2}(\theta) = \bar n^r_{2} (0) \exp\left[-g_{rr}(\bar n_1) t_\text{int} \sin^2\theta\right]$ 
where $t_\text{int}=\SI{14}{\micro \s}$ is the total interaction time and $g_{rr}(\bar{n}_1)\sin^2\theta = \gamma_{rr}$.
In the \figref{fig:MW_pulses}{d} we see that measured interaction strength values follow the expression 
$g_{rr}(\bar n_1) = g_1 \bar n_1$.
From the fitted linear function, the decay value for the single Rydberg excitation is $g_1 = \SI{0.63(0.08)}{\kilo \hertz}$
which is very close to the predicted theoretical value 
$g_{1,\text{theory}} = 2Q/\mathcal V= \SI{0.56(0.13)}{\kilo \hertz} $
of the decay per single Rydberg excitation
as calculated in the two-particle interaction model of \cite{kurzyna2025microwavefieldquantummetrologyinherent}.  
We benchmarked the controlled attenuation to reduce the intensity of the signal by 50 \% in $\SI{14}{\micro \s}$ with about 100 Rydberg excitations, which corresponds to excitation density $n_{\rho} = \SI{196}{\per \cubic \milli \meter} $

\begin{figure*}
    \centering
    \includegraphics[width=\linewidth]{ 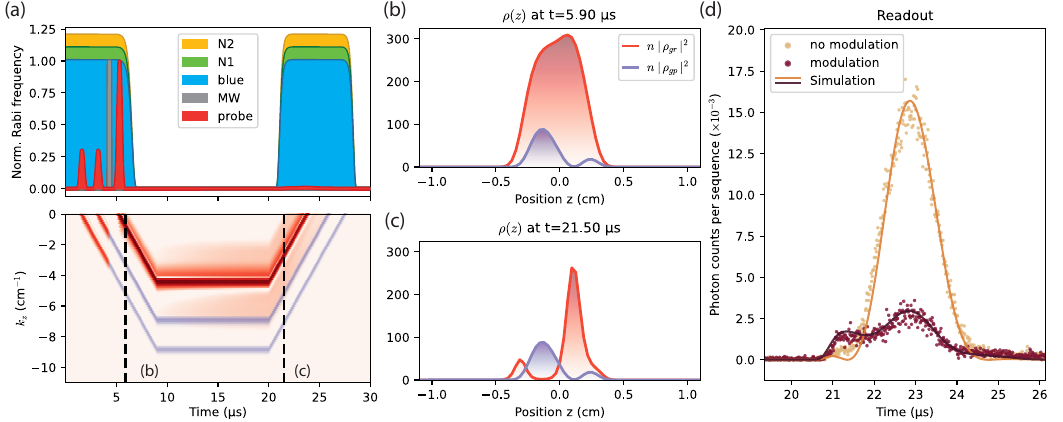}
    \caption{Interaction-induced diffraction.
    \subfig{a} Sequence of operations (upper plot) 
    and the simulation of the memory contents, i.e., stored spinwaves 
    as a colormap of the Fourier transform along the $z$-axis (lower plot). 
    Red map represents coherence $\rho_{gs}$, and the purple map corresponds to $\rho_{gp}$. 
    \subfig{b},\subfig{c} Simulated amplitudes of the stored atomic coherences as a function of the position along the atomic ensemble right after the storage of the signal and right before the readout.
    During the storage (from $t = \qty{6}{\us}$ to $t=\qty{20}{\us}$), the dipolar interactions between the Rydberg atoms disturb the spinwave coherence in regions where both $\ket{s}$ and $\ket{p}$ overlap, reducing the readout from those positions.
    The amplitude modulation pattern is imprinted onto the spinwave in $\ket{s}$, resulting in the emergence of sidebands in the readout signal.
    \subfig{d} Points representing measured readout with and without the gating pulses. Solid lines are the simulated readout.
}
    \label{fig:diffraction}
\end{figure*}

\subsection{Controlled diffraction with the Rydberg spinwave} 

The combination of the GEM multiplexing capabilities and strong Rydberg interactions allows for the control of the spatial distribution of the Rydberg excitations in the different states. By switching the position-dependent interactions, we can engineer the controlled diffraction of the signal on the interference pattern created with the previously stored spinwave.

To demonstrate the controlled diffraction, we use a composite gating pulse which is a pair of Gaussians with duration $\tau_{\text{CTRL}} = \qty{0.2}{\us}$, each separated by $T=\qty{2.25}{\us}$. The sequence is presented in \figref{fig:diffraction}{a}, showing the light and matter domains in vertical and horizontal planes, respectively. The gate pulses are initially stored in the Rydberg state $\ket{s}$. The GEM mapping of the gating pulses creates a Rydberg coherence with a spatial interference pattern in the atomic ensemble along the $z$-axis as $\rho_{gs}(z)\propto \cos({\beta z T}/{2})$. Later, around $t=\qty{4.3}{\us}$, the coherence is transferred to the nearby Rydberg state $\ket{p}$ with MW $\pi$-pulse. Next, after $\qty{2.1}{\us}$ the signal pulse is stored in the state $\ket{s}$. The GEM gradient is then reversed, and $\Delta t = \qty{14}{\us}$ after the write-in, the controls are re-enabled, and the read-out is captured. We input the experimental settings to the simulation, including a small relative phase between these pulses, which results from a nonzero four-photon detuning. The simulated coherences immediately after signal storage are shown in \figref{fig:diffraction}{b}.

Due to the dipolar coupling between the Rydberg states $\ket{s}$ and $\ket{p}$,
the spatial profile of $n(z)|\rho_{gp}(z)|^2$ interacts with the stored signal coherence $\rho_{gs}(z)$. 
The resulting large phase imprinted on the signal coherence at the positions where $\ket{p}$ was populated
disables the readout from those positions and effectively acts as an amplitude modulation profile.
This is modeled as the decay of the $\rho_{gs}(z)$ coherence proportional to the spatially varying density $n(z)|\rho_{gp}(z)|^2$. 
The simulation results for $\rho_{gs}(z)$ and $\rho_{gp}(z)$ after the interaction at time $t = \qty{21.5}{\us}$ are depicted in \figref{fig:diffraction}{c}. The hole in $\rho_{gs}(z)$ around $z\simeq-2$mm 
corresponds to the local maximum of $\ket{p}$ population.

The modulated signal coherence is then read out from the memory by reversing the magnetic field gradient. The simulation of the resulting readout is presented in \figref{fig:diffraction}{d}
for the Rydberg spinwave in the absence of gating pulses (red curve) and in their presence (orange curve). The measured photon counts are binned; each bin has a width of \qty{12.5}{\ns}. The solid lines are the simulated readout. The simulated results are scaled and multiplied by an exponential with $\tau_e=\qty{3}{\us}$, which reflects the additional observed decay during readout.
The reversed mapping during GEM readout translates the wavevector components of $\tilde\rho_{gs}(k_z)$ to the temporal envelope of the signal pulse $\Omega_p(t)$. 
Therefore, the grating-like modulation along $z$ is transformed to the diffraction orders in the temporal profile of the readout, 
making them detectable as temporal sidebands.

\section{Discussion}
We have demonstrated the novel four-photon adressing and interfacing in a \'{N}-letter shape configuration that allows for the instantaneous generation of the Rydberg spinwaves with near-zero wavevector.
The memory lifetime was increased by almost an order of magnitude, leading to the stored information decay over $\qty{20}{\micro \s}$ being comparable to the radiative decay of the excited state.
This prolonged the lifetime of the spinwave and enabled the use of GEM multimode access to the memory allowing for temporal and spectral multiplexing of the signal.
The robustness of this method comes from the utilization of virtual states in the D2-line. The transitions to those states are enabled with easily accessible high-power lasers. 

We leveraged microwave transitions between neighbouring Rydberg $\ket s$ and $\ket p$ states to transfer the created coherence. 
We observed and quantified the strong interactions between said states, obtaining good agreement of measured values with theoretical predictions. 
In particular, we demonstrated that one pulse stored in GEM can controllably attenuate another pulse. 
About 100 control Rydberg excitations corresponding to an excitation density of $\SI{196}{\per \cubic \milli \meter} $ can produce 50\% attenuation of the signal in \qty{14}{\us}.

The frequency-to-position mapping provided in GEM enabled spatial control of Rydberg excitations whose interactions were tuned by microwave fields. 
We demonstrated controllable diffraction of the stored optical pulse by first storing two gating pulses in a Rydberg spinwave and then transferring them to a different, strongly interacting Rydberg state.

The current parameters of the setup allow for the demonstration of the presented scheme but can still be improved. A higher density of the atomic ensemble would allow for higher storage and retrieval efficiency. 
The larger bandwidth of the memory could be achieved by increasing the value of the magnetic field gradient or by elongating the atomic cloud. Larger bandwidth would be essential for storing a higher number of modes in the memory, allowing for more precise control of the position of stored information.

Similar protocols can be applied to create spinwaves insensitive to thermal decoherence in all ladder configurations of Rydberg memories in alkali atoms.

We envisage the development of the presented scheme in two directions. Firstly, long-lived Rydberg spinwaves can now be successfully applied in longer, ultrasensitive microwave field detection. 
Secondly, spatial control of the interactions introduces many prospects for imaging individual Rydberg excitations \cite{Du:25}. 
By creating a single excitation in the same position in the atomic ensemble for every experimental realization, many frames can be averaged to obtain the image instead of single-shot imaging. 
Moreover, the storage of the Rydberg excitation in GEM paves the way for the tomography of the propagation of an individual Rydberg polariton, without the use of spatial homodyne detection on the camera.

\section*{Data Availability}
Data that supports this study has been deposited at \cite{OJTQQJ_2026}.

\section*{Author contribution}
BN and SK set up the experimental system, measured the data, and wrote the first version of the manuscript. BK helped with rigorous derivations of the theoretical model and helped refine the article. WW and MP proposed, supervised the research, and helped at various stages of the experiment. All authors reviewed the manuscript.
\begin{acknowledgments}
The “Quantum Optical Technologies” (FENG.02.01-IP.05-0017/23) project is carried out within the Measure 2.1 International Research Agendas programme of the Foundation for Polish Science, co-financed by the European Union under the European Funds for Smart Economy 2021--2027 (FENG). This research was funded in whole or in part by the National Science Centre, Poland, grant no. 2024/53/B/ST2/04040. Publication co-financed from the state budget funds (Poland), awarded by the Minister of Science under the “Perły Nauki II” program, project No. PN/02/0027/2023, co-financing amount PLN 239,998.00, total project value PLN 239,998.00.
\end{acknowledgments}

\bibliographystyle{quantum}
\bibliography{refs}

\clearpage
\appendix
\section{Atom-light interface}
\label{sec:theory}
To model the system, we utilize an atomic semi-classical Hamiltonian. We take into consideration 5 levels ($g, e, h, f, s$) as building blocks of the excitation scheme, as depicted in \figref{fig:levels_scheme_lifetime}{a}. For now, we omit the highest Rydberg state $\ket{p}$. Fields enter the Hamiltonian by defined Rabi frequency: $\Omega = \frac{d E}{\hbar}$, where $d$ is the dipole moment magnitude on the selected transition and $E$ is the amplitude of the electric field driving the transition. To simplify the calculations, we move to the interaction frame and perform the rotating wave approximation, resulting in a time-independent, effective Hamiltonian of the following form: 
\begingroup
\setlength{\arraycolsep}{1pt} 
\begin{equation}
        H = -\hbar 
    \scalebox{0.95}{$
    \begin{pmatrix}
        0 & \frac{\Omega^*_{p}}{2} & 0 & 0 & 0 \\
         \frac{\Omega_{p}}{2} & - \tDelta &  \frac{\Omega_{N1}}{2} & 0 & 0 \\
        0 &  \frac{\Omega^*_{N1}}{2} & - \Delta_{N1} & \frac{\Omega^*_{N2}}{2} & 0 \\
        0 & 0 &  \frac{\Omega_{N2}}{2} & -\tDelta_{N2}&  \frac{\Omega^*_{c}}{2} \\
        0 & 0 & 0 &  \frac{\Omega_{c}}{2} & \delta + \iu \frac{\gamma_{rr}}{2} \\
        
    \end{pmatrix}$},
\end{equation}
\endgroup
where we account for system decoherence by introducing the decays of states $\ket{e}$ and $\ket{f}$ with rates $\gamma_{e}$ and $\gamma_{f}$ respectively, in its diagonal.
For the clarity of the equations, we use complex detunings $\tDelta =\Delta -i \gamma_e/2$ and $\tDelta_{N2} = \Delta_{N2} -i \gamma_f/2$. 
The $\delta$ is taken at a single $z$ position, but is further generalized to position-dependent detuning. 
Additionally, we account for the interaction-induced decay $\gamma_{rr}$ of the state $\ket{s}$, explained in detail in Sec.~\ref{sec:Decay}.
Repopulation of ground states occurs at rate $\gamma_{e} \rho_{ee}$ for state $\ket{g}$ and $\rho_{ee} \gamma_{e} + \rho_{ff} \gamma_{f}$ for $\ket{h}$. Those are elements of the repopulation matrix $Y$. Together, the full master equation is given by
\begin{equation}
    \dot \rho = -i H\rho + i \rho H^\dagger + Y. 
\end{equation}

To further simplify the equations, we approximate them to the first order with respect to the probe Rabi frequency $\Omega_{p}$ and use adiabatic elimination. 
We assume all atoms are in the ground state $\rho_{gg} =1$ and $\rho_{ii}=0$ for other levels. 
We also solve for coherences assuming they follow steady state $\dot{\rho}_{ij} = 0$
for $i\ne j$ except $\rho_{gs}$.  
Ultimately, in the low-excitation limit, the introduced $\gamma_{rr}$ serves only as a dephasing factor of the coherence $\rho_{gs}$ and does not account for ground state repopulation. 
This way we can express said coherences as functions of $\rho_{gs}$ and insert their adiabatic values 
into the equation for $\dot\rho_{gs}$.

The equation for propagation of probe pulse comes from considering electric field evolution in the slowly varying envelope approximation, 
and substituting atomic polarization as $n(z)d_{ge}\rho_{ge}$ using $\rho_{ge}$ obtained from adiabatic elimination, where $n(z)$ is atomic density profile.
This way we recover Eqs.~\eqref{MBEs}:

\begin{widetext}
\begin{subequations}
\begin{equation}
\dd{\Omega_p}{z} = \frac{n(z)}{N} \, \frac{\text{OD}}{2}\Biggl[
\overbrace{\frac{\iu \gamma_e\Omega_{C} \Omega_{N1} \Omega_{N2}}{8\tDelta \Delta_{N1} \tDelta_{N2} - 2\tDelta_{N2} |\Omega_{N1}|^2 - 2\tDelta   |\Omega_{N2}|^2}}^{C_{\Omega,\rho}}\rho_{gs}
+ \overbrace{\frac{\iu \gamma_e\left(4\Delta_{N1}   \tDelta_{N2} - |\Omega_{N2}|^2\right)}{8\tDelta   \Delta_{N1}   \tDelta_{N2} - 2\tDelta_{N2}   |\Omega_{N1}|^2 - 2\tDelta   |\Omega_{N2}|^2}}^{C_{\Omega,\Omega}} \Omega_p
\Biggr]
\end{equation}
\begin{multline}
\dd{\rho_{gs}}{t} = \underbrace{\frac{\iu \Omega_{C}   \Omega_{N1}   \Omega_{N2}}{16\tDelta   \Delta_{N1}   \tDelta_{N2} - 4\left(\tDelta_{N2}   |\Omega_{N1}|^2 + \tDelta   |\Omega_{N2}|^2\right)}}_{C_{\rho,\Omega}} \Omega_p\\
+  \Biggl( \underbrace{\iu\frac{|\Omega_{C}|^2   \left(4\tDelta   \Delta_{N1} - |\Omega_{N1}|^2\right)}{16\tDelta   \Delta_{N1}   \tDelta_{N2} - 4\left(\tDelta_{N2}   |\Omega_{N1}|^2 + \tDelta |\Omega_{N2}|^2\right)}}_{C_{\rho,\rho}} + \iu\delta(z) - \frac{\gamma_{rr}}{2} \Biggr)\rho_{gs} 
\end{multline}
\end{subequations}
\end{widetext}

\section{Rydberg interactions}
\label{sec:Decay}
Microscopically, each atom in $\ket{p}$ imprints a large phase on the signal coherence $\rho_{gs}$ in its vicinity, disallowing for the readout from those positions \cite{kurzyna2025microwavefieldquantummetrologyinherent}.
We quantify the interaction strength by calculating the effective decay $\gamma_{rr}$ of the coherence after the interactions during the storage in the memory: 
\begin{equation}
    \gamma_{rr} = \frac{2Q \bar n_p}{\mathcal{V}} \; \text{and} \; Q = \frac{4 \pi^2}{9 \sqrt{3}}  \frac{C_3}{\hbar} 
    \label{gamma_{rr}=Qn_V}
\end{equation}
where the population of the Rydberg atoms in the state $\ket{p}$ is $\bar n_p = N|\rho_{gp}|^2$ and $\mathcal{V}$ is the volume of the ensemble.
We can understand the factor $\bar n_p Q t$ as the excluded volume in the ensemble due to the interactions that every atom in the state $\ket{p}$ generates.
For the states $59S$ and $59P$ chosen in the experiment, the dipolar interaction strength is described with $\frac{C_3}{2\pi\hbar} = \SI{4.111}{\GHz \: \um^3}$ \cite{SIBALIC2017319}.

By dividing the cloud into sections and considering the decay in each of them independently, we find the position-dependent decay rate. 
The ratio $\bar n_p/\mathcal{V}$ appearing in \eqref{gamma_{rr}=Qn_V} is naturally replaced with $n(z) \rho_{pp}$ where 
$\rho_{pp} \approx |\rho_{gp}|^2$ is the probability of atoms being in state $\ket{p}$.
This way we find that the position-dependent decay rate due to interactions equals $\gamma_{rr}=2n(z)Q|\rho_{gp}|^2$ included in \eqref{MBEs}.
\section{Simulation}
Simulating the propagation equations was performed using XMDS2\cite{DENNIS2013201}.

\end{document}